\documentclass[prd,showpacs,superscriptaddress,nofootinbib,floatfix,11pt]{revtex4-2}

\usepackage{geometry}
\usepackage{amsmath, amssymb}
\usepackage{mathtools}
\usepackage{color}
\usepackage{slashed}
\usepackage{bbm}
\usepackage{graphicx}
\usepackage[caption=false]{subfig}
\usepackage[colorlinks=true,
breaklinks=true,
urlcolor=magenta,
citecolor=blue]{hyperref}
\usepackage{orcidlink}
\usepackage[normalem]{ulem}

\newcommand{\itp}{\affiliation{CAS Key Laboratory of Theoretical Physics, Institute of Theoretical Physics,\\ Chinese Academy of Sciences, Beijing 100190, China}}
\newcommand{\ucas}{\affiliation{School of Physical Sciences, University of Chinese Academy of Sciences, Beijing 100049, China}}
\newcommand{\peng}{\affiliation{Peng Huanwu Collaborative Center for Research and Education,\\ Beihang University, Beijing 100191, China}}
\newcommand{\hiskp}{\affiliation{Helmholtz-Institut f\"{u}r Strahlen- und Kernphysik and Bethe Center for Theoretical Physics,\\ Universit\"{a}t Bonn, D-53115 Bonn, Germany}}
\newcommand{\fzj}{\affiliation{Institute for Advanced Simulation and Institut f\"{u}r Kernphysik, Forschungszentrum J\"{u}lich, D-52425 J\"{u}lich, Germany}}
\newcommand{\tbilisi}{\affiliation{Tbilisi State University, 0186 Tbilisi, Georgia}}

\begin{document}

\title{Pion axioproduction revisited}
\author{Cheng-Cheng Li}\email{lichengcheng@itp.ac.cn}\itp\ucas
\author{Tao-Ran Hu}\email{hutaoran21@mails.ucas.ac.cn}\ucas
\author{Feng-Kun Guo\orcidlink{0000-0002-2919-2064}}\email{fkguo@itp.ac.cn}\itp\ucas\peng
\author{Ulf-G. Mei{\ss}ner\orcidlink{0000-0003-1254-442X}}\email{meissner@hiskp.uni-bonn.de}\hiskp\fzj\tbilisi


\begin{abstract}
In this work, we extend the analysis of the pion axioproduction, $aN \to \pi{N}$,
to include the impact of the Roper resonance $N^*(1440)$ together with the previously studied $\Delta(1232)$ resonance.
Our theoretical framework is a chiral Lagrangian approach with explicit resonance fields to account for their respective impacts.
We find that the $N^*(1440)$ also leads to an enhancement of the cross section within its energy range for various axion models.
\end{abstract}

\maketitle

\newpage 

\section{Introduction}
The axion is a well-motivated paradigm for physics beyond the Standard Model, simultaneously providing
a solution to the strong CP problem~\cite{Peccei:1977hh, Peccei:1977ur, Weinberg:1977ma, Wilczek:1977pj}
and serving as a potential candidate for dark matter~\cite{Preskill:1982cy, Abbott:1982af, Dine:1982ah}.
The original (``visible'') Peccei–Quinn–Weinberg–Wilczek (PQWW) axion with a decay constant $f_a$ at the
electroweak scale (or equivalently a mass $m_a \approx 5.7\times(10^6\,\text{GeV}/f_a)\,\text{eV}$ in
the $\text{keV}/\text{MeV}$ region) was quickly ruled out by experiments on astrophysical grounds
(axion emission from the sun and red giants)~\cite{Dicus:1978fp, Dicus:1979ch}. Thus, the ``invisible”
axion was introduced with an extraordinarily large decay constant traditionally estimated to be
$10^{9}\,\text{GeV} \lesssim f_a \lesssim 10^{12}\,\text{GeV}$ (corresponding to an axion mass between a
few $\mu\text{eV}$ and $0.1\,\text{eV}$)~\cite{Kim:2008hd}, such as the Kim–Shifman–Vainstein–Zakharov
(KSVZ) axion model~\cite{Kim:1979if, Shifman:1979if} or the Dine–Fischler–Srednicki–Zhitnitsky (DFSZ)
axion model~\cite{Dine:1981rt, Zhitnitsky:1980tq}.

Astrophysical observations can place stringent bounds on the properties of the axion. For instance, a
core-collapse supernova (SN), e.g. SN~1987A, can emit axions in addition to neutrinos as an extra
cooling mechanism of the associated neutron star. Consequently, the suppression of the neutrino
luminosity due to axion emission would discernibly alter the observed neutrino events to provide
stringent bounds on the axion-nucleon couplings~\cite{Raffelt:1990yz, Engel:1990zd}.

Recently, Carenza et al.~\cite{Carenza:2020cis} revisited the axion emissivity due to the pion-induced
process $\pi^-p \to an$ and pointed out that SNe can emit axions with energies up to $500\,\text{MeV}$,
which in turn can produce pions in water Cherenkov detectors via the $aN \to \pi{N}$ process.
At these energies, an enhanced cross section of the pion axioproduction can be expected due to the
intermediate resonances. Note that we use the term axioproduction in analogy with terms like pion
electro- or photoproduction. Pion axioproduction hence means pion production induced by axions.
In Ref.~\cite{Vonk:2022tho}, by conducting a study on the $P_{33}$ partial wave cross section of this process,
the authors confirmed the existence of such an enhancement in the Delta resonance $\Delta(1232)$ region, which
can be accessed when the axion energy $E_a \simeq 200\text{--}300\,\text{MeV}$.
They also pointed out that the $\Delta(1232)$ contribution to the $aN \to \pi{N}$ process breaks isospin symmetry with the amplitude proportional to $(m_d-m_u)/(m_d+m_u)$.
Thus, the enhancement of the cross section of pion axioproduction estimated in Ref.~\cite{Carenza:2020cis} is reduced by $1$ to $5$ orders of magnitude.

In this work, we take a step further by considering also the effects of the Roper resonance $N^*(1440)$ on this process, whose contribution conserves isospin.
This is motivated by the simple observation that the invariant mass of the initial $aN$ system falls within the $N^*(1440)$ region
when the axion energy $E_a$ is approximately in the range of $400\text{--}500\,\text{MeV}$, which is on the right shoulder of the bump of the SN emitted axion number spectrum from the $\pi^-{p}\to an$ process derived in Ref.~\cite{Carenza:2020cis}.
Furthermore, the $N^*(1440)$ decays into $\pi{N}$ with a large branching fraction of $(55\text{--}75)\%$~\cite{ParticleDataGroup:2022pth}. Hence, it is imperative for us to consider its impact.
While the $N^*(1440)$ does not couple as strongly as the $\Delta(1232)$ to the pion-nucleon system,
the fact that the pion axioproduction via the $N^*(1440)$ is isopsin-conserving counteracts this suppression.
In fact, we will demonstrate that the $N^*(1440)$ also leads to an enhancement in the cross section, and further implications for experimental detection of the axion are discussed.

We employ a chiral Lagrangian framework with explicit resonance fields.
The chiral Lagrangian is the leading order (LO) one in chiral perturbation theory (ChPT), which
is a low-energy effective theory of quantum chromodynamics (QCD)~\cite{Weinberg:1978kz, Gasser:1983yg}.
In ChPT,  the pions and nucleons, rather than the more fundamental quarks and gluons, are treated as the effective degrees
of freedom, while the axion can be incorporated through external sources. Additionally, we explicitly introduce
resonance fields, namely the $\Delta (1232)$ and $N^* (1440)$ fields, following Ref.~\cite{Meissner:1999vr}, to account for their effects. This framework enables us
to draw upon established knowledge of hadronic processes while at the same time preserve the consequences of the spontaneously broken
chiral symmetry of QCD.

The outline of this paper is as follows: In Sec.~\ref{sec: kinematics}, we collect the necessary kinematics concerning pion axioproduction.
In Sec.~\ref{sec: incorporation of axion}, we outline the main steps for incorporating the axion into ChPT.
The Lagrangians describing the axion-nucleon and axion-resonance interactions are collected in Sec.~\ref{sec: evaluation of Feynman diagrams} where we also evaluate their contributions to the scattering amplitude.
Subsequently, we assemble these contributions and proceed to analyze the obtained results in Sec.~\ref{sec: results}.

\section{Kinematics} \label{sec: kinematics}
In this section, we give a short discussion of the general isospin structure of the scattering amplitude of pion
axioproduction and its partial wave decomposition, following closely Ref.~\cite{Vonk:2022tho}. This serves
to set our notation and to keep the manuscript self–contained. The process under consideration is
\begin{equation}
a(q) + N(p) \to \pi^b(q^\prime) + N(p^\prime),
\end{equation}
where $a$ denotes an axion, $N$ a nucleon, either proton or neutron, and $\pi^b$ a pion with the Cartesian
isospin index $b$. As usual, we define the Lorentz-invariant Mandelstam variables:
\begin{equation}
s = {(p+q)}^2, \qquad t = {(p-p^\prime)}^2, \qquad u = {(p-q^\prime)}^2.
\end{equation}
These invariants fulfill the on-shell relation,
\begin{equation} \label{eq: Mandelstam identity}
s+t+u = 2m_N^2 + m_a^2 + M_\pi^2,
\end{equation}
which can be used to eliminate one of the three variables, which we choose to be $u$. In what follows,
we will take the isospin-averaged nucleon mass $m_N = 938.92\,\text{MeV}$ and the isospin-averaged
pion mass $M_\pi = 138.03\,\text{MeV}$. Throughout this paper, we use the center-of-momentum (c.m.)
frame, for which the three-momenta obey the relation $\mathbf{p}+\mathbf{q}=\mathbf{p^\prime}+\mathbf{q^\prime}=0$.
Using the well-known K\"all\'{e}n function,
\begin{equation}
\lambda(a,b,c) = a^2+b^2+c^2-2ab-2ac-2bc,
\end{equation}
one has
\begin{align}
\begin{split}
|\mathbf{p}|=|\mathbf{q}| &= \frac{\sqrt{\lambda(s,m_N^2,m_a^2)}}{2\sqrt{s}}, \\
|\mathbf{p^\prime}|=|\mathbf{q^\prime}| &= \frac{\sqrt{\lambda(s,m_N^2,M_\pi^2)}}{2\sqrt{s}},
\end{split}
\end{align}
and the c.m. energies of the incoming and outgoing nucleons can be written as
\begin{equation}
E_\mathbf{p} = \frac{s+m_N^2-m_a^2}{2\sqrt{s}},\qquad E_\mathbf{p^\prime} = \frac{s+m_N^2-M_\pi^2}{2\sqrt{s}}.
\end{equation}
Moreover, setting $z=\cos{\theta}$, where $\theta$ is the c.m. scattering angle, we have
\begin{equation}
\mathbf{p}\cdot\mathbf{p^\prime} = |\mathbf{p}||\mathbf{p^\prime}|z,
\end{equation}
so we can reexpress the second Mandelstam variable $t$ as
\begin{equation}
t = 2\left(m_N^2 - E_\mathbf{p}E_\mathbf{p^\prime} + |\mathbf{p}||\mathbf{p^\prime}|z\right).
\end{equation}

In the following, we consider the scattering amplitude $T^{b}_{aN \to \pi{N}}$. According to the
isospin structure, it can be parameterized as
\begin{equation}
T^{b}_{aN \to \pi{N}} = \frac{1}{2}\{\tau^b,\tau^3\}T^+ + \frac{1}{2}[\tau^b,\tau^3]T^- + \tau^bT^3,
\end{equation}
which is similar to the case of $\pi{N}$ elastic scattering with isospin violation, see, e.g.
Ref.~\cite{Fettes:2000vm}. Any of the four possible scattering amplitudes can then be expressed
in terms of the three amplitudes $T^{+/-/3}$:
\begin{align}
\begin{split} \label{eq: T^{+/-/3}}
T_{ap \to \pi^0p} & = T^+ + T^3, \\ 
T_{an \to \pi^0n} & = T^+ - T^3, \\
T_{ap \to \pi^+n} & =  \sqrt{2}\left(T^- + T^3\right), \\
T_{an \to \pi^-p} & = -\sqrt{2}\left(T^- - T^3\right).
\end{split}
\end{align}
Furthermore, according to the Lorentz structure, each of the three amplitudes $T^{+/-/3}$ can be
decomposed as (the superscripts are suppressed for simplicity)
\begin{equation}
T(s,t;\lambda^\prime,\lambda) = \bar{u}(p^\prime,\lambda^\prime) \left\{A(s,t) + B(s,t)\frac{1}{2}\left(\slashed{q}
+\slashed{q}^\prime\right)\right\} u(p,\lambda),
\end{equation}
where $\lambda^{(\prime)}$, appearing in the Dirac spinor, denotes the helicity of the incoming (outgoing) nucleon.
The partial wave amplitudes $T^{l\pm}(s)$, where $l$ refers to the orbital angular momentum and the
superscript $\pm$ to the total angular momentum $j = l\pm1/2$, are given in terms of the functions $A(s,t)$
and $B(s,t)$ via
\begin{align}
\begin{split} \label{eq: partial wave projection}
T^{l\pm}(s)
=&\, \frac{\sqrt{E_\mathbf{p}+m_N}\sqrt{E_\mathbf{p^\prime}+m_N}}{2}\left\{A^l(s) + \left(\sqrt{s}-m_N\right)B^l(s)\right\} \\
&+ \frac{\sqrt{E_\mathbf{p}-m_N}\sqrt{E_\mathbf{p^\prime}-m_N}}{2}\left\{-A^{l\pm1}(s) + \left(\sqrt{s}+m_N\right)B^{l\pm1}(s)\right\},
\end{split}
\end{align}
where
\begin{align}
\begin{split}
A^l(s) &= \int_{-1}^{+1}{\text{d}z}\:{A(s,t(s,z))P_l(z)}, \\
B^l(s) &= \int_{-1}^{+1}{\text{d}z}\:{B(s,t(s,z))P_l(z)}.
\end{split}
\end{align}
The total cross section can
be expanded in terms of the partial wave cross sections as~\cite{Jacob:1959at}
\begin{equation}
\sigma = \sum_{l\pm}{\sigma^{l\pm}},
\end{equation}
where
\begin{equation} \label{eq: partial wave cross section}
\sigma^{l\pm} = \frac{1}{32\pi{s}}\frac{|\mathbf{p^\prime}|}{|\mathbf{p}|}(2l\pm1+1){|T^{l\pm}|}^2.
\end{equation}

In this work, we perform the calculation of the $S_1$, $P_1$ and $P_3$ partial wave cross sections while neglecting the higher ones with $l \geq 2$, as those are suppressed in the energy region under consideration.
Throughout, we make use of the notation $l_{2j}$, with $l=S,P,D,\cdots$ the orbital angular momentum, and $j$ the total angular momentum.
Each of the partial waves contains both the isospin-conserving $(I=1/2)$ and isospin-breaking ($I=3/2$) contributions.
That is, $S_1$ refers to the sum of both $S_{11}$ and $S_{31}$ (in the usual $l_{2I,2j}$ notation) partial waves, and so on.
Since the $\Delta$ is a spin-$\tfrac{3}{2}$, positive-parity resonance and the Roper is a spin-$\tfrac{1}{2}$, positive-parity resonance, it is reasonable to expect an enhancement in the $P_3$ and $P_1$ partial wave cross sections in the energy region of the $\Delta$ and Roper resonances, respectively.
Of course, we are well aware that the Roper does not show up as a bump in the pion-nucleon cross section and the $P_{11}$ phase shift crosses $90^\circ$ at an energy higher than $1.44\,\text{GeV}$.
In case of pion axioproduction, matters can be different as the background is much suppressed.

\section{Incorporation of the axion into ChPT} \label{sec: incorporation of axion}
In this section, we give a brief presentation of how the axion can be incorporated into ChPT.
For a more detailed discussion, we refer to Refs.~\cite{Georgi:1986df, Vonk:2020zfh, Vonk:2021sit, Meissner:2022cbi}.
Consider the general QCD Lagrangian with axion below the Peccei-Quinn (PQ) symmetry breaking scale
\begin{equation}
\mathcal{L}_{\text{QCD}} = \mathcal{L}_{\text{QCD},0} + \frac{a}{f_a}{\left(\frac{g}{4\pi}\right)}^2\text{Tr}
\left[G_{\mu\nu}\tilde{G}^{\mu\nu}\right] + \bar{q}\gamma^\mu\gamma_5\frac{\partial_\mu a}{2f_a}\mathcal{X}_qq,
\end{equation}
where $q = {(u,d,s,c,b,t)}^\text{T}$ collects the quark fields, $a$ refers to the axion field, and $f_a$ is
the axion decay constant. Depending on the underlying axion model, the coupling constants of the axion-quark
interactions in the matrix $\mathcal{X}_q=\operatorname{diag}(X_q)$ are given by
\begin{align}
\begin{split} \label{eq: X_q}
X_q^{\text{KSVZ}} &= 0, \\
X_{u,c,t}^{\text{DFSZ}} &= \frac{1}{3}\frac{x^{-1}}{x+x^{-1}} = \frac{1}{3}\sin^2{\beta}, \\
X_{d,s,b}^{\text{DFSZ}} &= \frac{1}{3}\frac{x}{x+x^{-1}} = \frac{1}{3}\cos^2{\beta},
\end{split}
\end{align}
for the KSVZ-type and DFSZ-type axion, respectively, where $x = \cot{\beta}$ is the ratio of the vacuum
expectation values (VEVs) of the two Higgs doublets in the latter model. After a suitable axial rotation of the
quark fields to remove the axion-gluon coupling term, the whole axion-quark interactions read
\begin{equation} \label{eq: axion-quark Lagrangian}
\mathcal{L}_{aq} = -(\bar{q}_L\mathcal{M}_aq_R+\text{h.c.}) + \bar{q}\gamma^\mu\gamma_5
\frac{\partial_\mu a}{2f_a}(\mathcal{X}_q-\mathcal{Q}_a)q,
\end{equation}
where
\begin{align}
\begin{split}
\mathcal{M}_a &= \exp{\left(i\frac{a}{f_a}\mathcal{Q}_a\right)}\mathcal{M}_q, \\
\mathcal{Q}_a &= \frac{\mathcal{M}_q^{-1}}{\langle\mathcal{M}_q^{-1}\rangle} \approx
\frac{1}{1+z+w}\text{diag}(1,z,w,0,0,0),
\end{split}
\end{align}
with $\mathcal{M}_q = \text{diag}\:\{m_q\}$ the quark mass matrix and $z = m_u/m_d$, $w = m_u/m_s$. We
take $z=0.485$ and $w=0.025$~\cite{FlavourLatticeAveragingGroupFLAG:2021npn}. 

It is from the interaction Lagrangian~\eqref{eq: axion-quark Lagrangian} that one has to determine
the axial-vector external sources $a_\mu$ (isovector) and $a_\mu^{(s)}$ (isoscalar) that enter ChPT.
In the $\text{SU}(2)$ case, this can be achieved by separating the $2$-dimensional flavor subspace of
the two lightest quarks from the rest and by decomposing the matrix $\mathcal{X}_q-\mathcal{Q}_a$ into a
traceless part and a part with non-vanishing trace, which results in
\begin{align}
\begin{split}
\mathcal{L}_{aq} = -(\bar{q}_L\mathcal{M}_aq_R+\text{h.c.}) &+ {\left(\bar{q}\gamma^\mu\gamma_5
\left(c_{u-d}\frac{\partial_\mu a}{2f_a}\tau_3+c_{u+d}\frac{\partial_\mu a}{2f_a}\mathbbm{1}\right)q
\right)}_{q={(u,d)}^\text{T}} \\
&+ \sum_{q=\{s,c,b,t\}}\left(\bar{q}\gamma^\mu\gamma_5c_q\frac{\partial_\mu a}{2f_a}q\right),
\end{split}
\end{align}
with
\begin{equation} \label{eq: axion couplings}
c_{u \pm d} = \frac{1}{2}\left(X_u \pm X_d - \frac{1 \pm z}{1+z+w}\right), \quad c_s = X_s - \frac{w}{1+z+w},
\quad c_{c,b,t} = X_{c,b,t}.
\end{equation}
Let $c_i$, $i=\{1,\cdots,5\}$, refer to the isoscalar couplings $\{u+d,s,c,b,t\}$, then one finds
\begin{equation}
a_\mu = c_{u-d}\frac{\partial_\mu a}{2f_a}\tau_3, \quad  a_{\mu,i}^{(s)} = c_i\frac{\partial_\mu a}{2f_a}\mathbbm{1}.
\end{equation}

With the usual $\text{SU}(2)$ matrix containing the three pions,
\begin{equation}
u = \sqrt{U} = \exp{\left(i\frac{\pi^a\tau_a}{2F}\right)},
\end{equation}
where $F$ is the pion decay constant in the chiral limit, for which we take the physical value
$F_\pi = 92.4\,\text{MeV}$ as the difference only amounts to effects of higher orders than
those considered here, one forms the following building blocks of ChPT:
\begin{align}
\begin{split}
D_\mu &= \partial_\mu + \Gamma_\mu, \text{ with }\Gamma_\mu = \frac{1}{2}[u^\dagger \partial_\mu u
+ u \partial_\mu u^\dagger - iu^\dagger a_\mu u + iu a_\mu u^\dagger], \\
u_\mu &= i[u^\dagger \partial_\mu u - u \partial_\mu u^\dagger - iu^\dagger a_\mu u - iu a_\mu u^\dagger], \\
u_{\mu,i} &= i[- iu^\dagger a_{\mu,i}^{(s)} u - iu a_{\mu,i}^{(s)} u^\dagger] = 2a_{\mu,i}^{(s)}.
\end{split}
\end{align}
Notice that, in principle, the axion can also enter ChPT through the building block,
\begin{equation}
\chi_\pm = u^\dagger \chi u^\dagger \pm u \chi^\dagger u, \text{ with }\chi = 2B\mathcal{M}_a,
\end{equation}
where $B$ is a constant related to the quark condensate $\Sigma = -\langle\bar{u}u\rangle$ in the chiral limit
via $B = \Sigma/F^2$. However, as this building block only appears in the interaction Lagrangians beyond leading
order, it will not be considered in what follows.

\section{Evaluation of the relevant Feynman diagrams} \label{sec: evaluation of Feynman diagrams}
In this section, we calculate several contributions to the scattering amplitude $T^{b}_{aN \to \pi{N}}$.
First, we consider the contact and nucleon-mediated diagrams, Fig.~\ref{fig: contact and nucleon}, arising
from the lowest order pion-nucleon Lagrangian. These diagrams start to contribute at $\mathcal{O}(q)$.
At $\mathcal{O}(q^2)$ and $\mathcal{O}(q^3)$, there are contributions arising from the pion-nucleon
Lagrangians beyond leading order. However, since axions have not been observed so far, some of the LECs of these higher-order interaction Lagrangians remain undetermined.
In our approach, such higher order contributions in the near-$\pi N$-threshold region are approximated by the explicit exchange 
of the $\Delta(1232)$ and of the $N^*(1440)$ resonances, see Figs.~\ref{fig: Delta} and \ref{fig: Roper}.
This is a sensible assumption as explained in Ref.~\cite{Bernard:1996gq}, where the dimension-two LECs were fixed from data and it was shown
that resonance saturation allows to explain these values.
The amplitudes in addition possess resonance poles in the energy region of interest that cannot be generated by a momentum expansion up to any finite order.
Finally, we will then consider the contributions from the pion rescattering loop diagram, Fig.~\ref{fig: rescatter}.
It is known that this type of contribution is most relevant for many processes at one-loop order,
with the notable exception of neutral pion photoproduction off protons or neutrons~\cite{Bernard:1991rt}.

\subsection{Contact term and graphs with an intermediate nucleon} \label{sec: contact and nucleon}
\begin{figure}
\centering
\subfloat[]{
\includegraphics[width=0.45\textwidth]{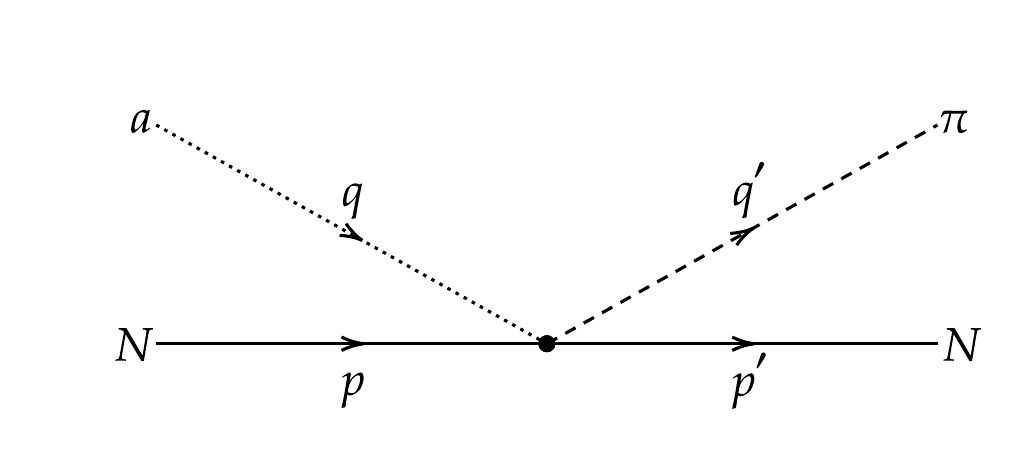}
\label{fig: contact}} \\
\subfloat[]{
\includegraphics[width=0.45\textwidth]{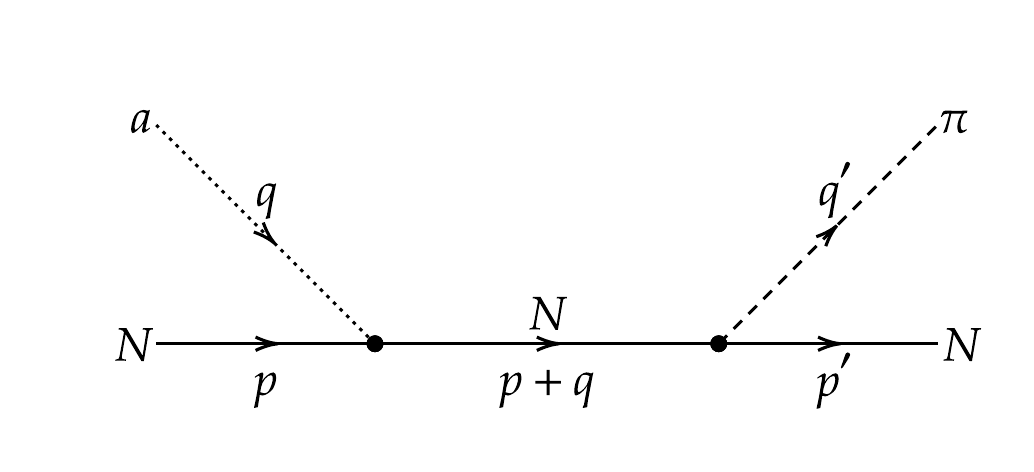}
\label{fig: Ns}}
\subfloat[]{
\includegraphics[width=0.45\textwidth]{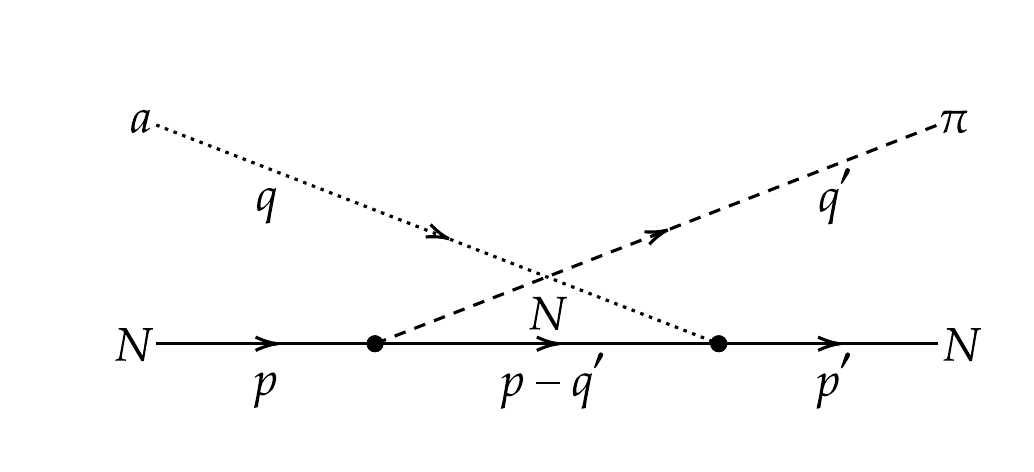}
\label{fig: Nu}}
\caption{Tree-level diagrams for $aN\to \pi N$ arising from the lowest order pion-nucleon Lagrangian:
(a) contact (Weinberg-Tomozawa) term, (b) $s$-channel $N$ exchange, and (c) $u$-channel $N$ exchange.}
\label{fig: contact and nucleon}
\end{figure}

In what follows, we only need the lowest order pion-nucleon Lagrangian, which is given by
\begin{equation} \label{eq: pion-nucleon Lagrangian}
\mathcal{L}_{\pi{N}}^{(1)} = \bar{\Psi}_N \left\{ i\slashed{D} - \mathring{m}_N
+ \frac{\mathring{g}_A}{2}\slashed{u}\gamma_5 +
\frac{\mathring{g}_0^i}{2}\slashed{u}_i\gamma_5 \right\} \Psi_N,
\end{equation}
where $\Psi_N = {(p,n)}^\text{T}$ is an isodoublet containing the proton and the neutron, $\mathring{m}_N$ is
the nucleon mass in the chiral limit, and $\mathring{g}_A$ and $\mathring{g}_0^i$'s are the axial-vector
isovector and isoscalar coupling constants, all also in the chiral limit. In
Eq.~\eqref{eq: pion-nucleon Lagrangian} and what follows, a summation over repeated $i$,
the index of isoscalar couplings, is implied. Again, to the order we are working, we can identify
these parameters with their physical values:
\begin{alignat}{2}
\mathring{g}_A &\to g_A &&= \Delta{u} - \Delta{d}, \notag\\
\mathring{g}_0^{u+d} &\to g_0^{u+d} &&= \Delta{u} + \Delta{d}, \\
\mathring{g}_0^q &\to g_0^q &&= \Delta{q}, \text{ for }q = s,c,b,t, \notag
\end{alignat}
where $s^\mu\Delta{q} = \langle{p}| \bar{q}\gamma^\mu\gamma_5q |p\rangle$, with $s^\mu$ the spin of the proton
and the superscript $\mu$ denoting the polarization direction. For these matrix elements, we take the
recent values from Ref.~\cite{FlavourLatticeAveragingGroupFLAG:2021npn},
\begin{align}
\Delta{u} =  0.847, \quad
\Delta{d} = -0.407, \quad
\Delta{s} = -0.035,
\end{align}
and ignore $\Delta{q}$ for $q=c,b,t$. The relevant diagrams from $\mathcal{L}_{\pi{N}}^{(1)}$
are depicted in Fig.~\ref{fig: contact and nucleon}.

The contact (Weinberg-Tomozawa) diagram, Fig.~\ref{fig: contact}, only gives a contribution to $B^-$:
\begin{equation}
B_\text{\ref{fig: contact}}^-(s,t) = \frac{c_{u-d}}{2f_a F_\pi}.
\end{equation}
For the $s$-channel nucleon-mediated diagram, Fig.~\ref{fig: Ns}, we find
\begin{align}
\begin{split} \label{eq: 1b}
&A_\text{\ref{fig: Ns}}^+(s,t) = A_\text{\ref{fig: Ns}}^-(s,t) = g_Ac_{u-d} \times A_N(s,t), \\
&A_\text{\ref{fig: Ns}}^3(s,t) = g_0^ic_i \times A_N(s,t), \\
&B_\text{\ref{fig: Ns}}^+(s,t) = B_\text{\ref{fig: Ns}}^-(s,t) = g_Ac_{u-d} \times B_N(s,t), \\
&B_\text{\ref{fig: Ns}}^3(s,t) = g_0^ic_i \times B_N(s,t),
\end{split}
\end{align}
where we have defined
\begin{align}
\begin{split}
A_N(s,t) &= \frac{g_Am_N}{2f_aF_\pi}, \\
B_N(s,t) &= -\frac{g_A}{4f_aF_\pi}\left(\frac{4m_N^2}{s-m_N^2}+1\right).
\end{split}
\end{align}
The $u$-channel diagram of Fig.~\ref{fig: Nu} can be obtained from the former  by crossing:
\begin{align} \label{eq: crossing}
A_\text{\ref{fig: Nu}}^+(s,t) &= +A_\text{\ref{fig: Ns}}^+(u,t),
&A_\text{\ref{fig: Nu}}^-(s,t) &= -A_\text{\ref{fig: Ns}}^-(u,t),
&A_\text{\ref{fig: Nu}}^3(s,t) &= +A_\text{\ref{fig: Ns}}^3(u,t), \notag\\
B_\text{\ref{fig: Nu}}^+(s,t) &= -B_\text{\ref{fig: Ns}}^+(u,t),
&B_\text{\ref{fig: Nu}}^-(s,t) &= +B_\text{\ref{fig: Ns}}^-(u,t),
&B_\text{\ref{fig: Nu}}^3(s,t) &= -B_\text{\ref{fig: Ns}}^3(u,t),
\end{align}
where $u$ needs to be understood as $u(s,t)$ via Eq.~\eqref{eq: Mandelstam identity}.

\subsection{Intermediate Delta and Roper resonances} \label{sec: Delta and Roper}
Next, we consider the exchange of the $\Delta$ resonance. The interactions of the $\Delta$ with
axions, pions and nucleons are given by the following effective Lagrangian, which is the leading
term of an appropriate chiral invariant Lagrangian~\cite{Hemmert:1997ye, Hacker:2005fh, Krebs:2008zb},
\begin{equation} \label{eq: Delta-pion-nucleon Lagrangian}
\mathcal{L}_{\Delta\pi{N}} = \frac{g}{2}\bar{\Delta}_\mu{T^{a\dagger}}(g^{\mu\nu}+z_0\gamma^\mu\gamma^\nu)\langle
\tau_au_\nu\rangle\Psi_N + \text{h.c.},
\end{equation}
where $\text{h.c.}$ stands for the Hermitian conjugate and $\langle\:\cdots\rangle$ denotes
the trace in flavor space. Here,
\begin{equation}
\Delta_\mu = \begin{pmatrix}
\Delta_\mu^{++} \\ \Delta_\mu^{+} \\ \Delta_\mu^{0} \\ \Delta_\mu^{-}
\end{pmatrix}
\end{equation}
collects the four $\Delta$ charge eigenstates, each of which is represented by a spin-$\tfrac{3}{2}$
vector-spinor field, and $T^a$'s are the isospin-$\tfrac{1}{2}\!\!\to\!\!\tfrac{3}{2}$ transition matrices.
The propagator of the $\Delta$ with four-momentum $p^\mu$ is then given by~\cite{Tang:1996sq, Krebs:2009bf}
\begin{equation}
-i\frac{\slashed{p}+m_\Delta}{p^2-m_\Delta^2}\left[ g^{\mu\nu} - \frac{1}{3}\gamma^\mu\gamma^\nu + \frac{1}{3m_\Delta}\left(p^\mu\gamma^\nu-\gamma^\mu p^\nu\right) - \frac{2}{3m_\Delta^2}p^\mu p^\nu \right],
\end{equation}
with $m_\Delta$ the mass of the $\Delta$, for which we take $m_\Delta = 1232\,\text{MeV}$.
For simplicity, we take here the Breit-Wigner rather than the pole mass, which is sufficient
for the accuracy of our calculation.
Moreover, the interaction Lagrangian~\eqref{eq: Delta-pion-nucleon Lagrangian} contains two coupling
constants $g = -1.366$ and $z_0 = -0.42$, whose values are taken from
Ref.~\cite{Meissner:1999vr}; see Fit~2 in Table~1 therein. 
Values from other fits in Ref.~\cite{Meissner:1999vr} will be used for an error estimate.
Notice that in the notation employed for the $\Delta$-pion-nucleon
Lagrangian in Ref.~\cite{Meissner:1999vr}, see Eq.~(3.5) therein, two coupling constants $g_{\Delta\pi{N}}$
and $Z$ appear. They are related with the ones in Eq.~\eqref{eq: Delta-pion-nucleon Lagrangian} via $g =
-\tfrac{F_\pi}{M_\pi}g_{\Delta\pi{N}}$ and $z_0 = -(Z+\tfrac{1}{2})$. We further note that the parameter $z_0$
can be eliminated, but this would just change the value of $g$. Here, we prefer to work with the
notation employed in Eq.~\eqref{eq: Delta-pion-nucleon Lagrangian}.

\begin{figure}
\centering
\subfloat[]{
\includegraphics[width=0.45\textwidth]{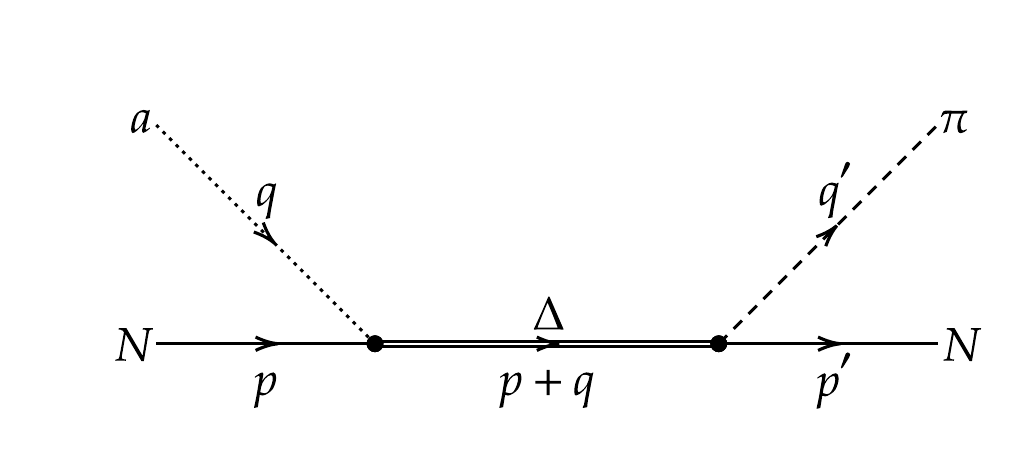}
\label{fig: Deltas}}
\subfloat[]{
\includegraphics[width=0.45\textwidth]{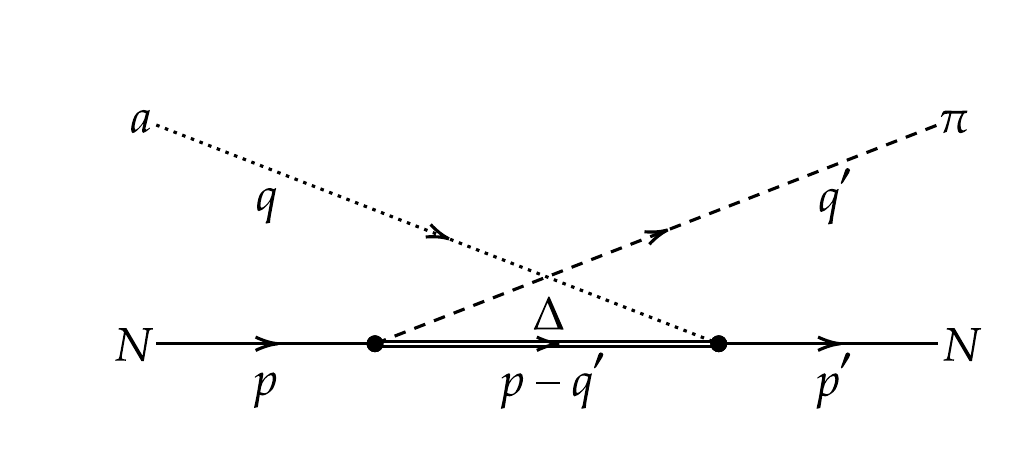}
\label{fig: Deltau}}
\caption{Diagrams  for $aN\to \pi{N}$ with the (a) $s$-channel and (b) $u$-channel exchange of the $\Delta$ resonance.}
\label{fig: Delta}
\end{figure}

For the contributions from the direct exchange of the $\Delta$, Fig.~\ref{fig: Deltas}, we find
\begin{align}
\begin{split} \label{eq: Deltas}
&A_\text{\ref{fig: Deltas}}^+(s,t) = -2A_\text{\ref{fig: Deltas}}^-(s,t) = A_\Delta(s,t), \\
&A_\text{\ref{fig: Deltas}}^3(s,t) = 0, \\
&B_\text{\ref{fig: Deltas}}^+(s,t) = -2B_\text{\ref{fig: Deltas}}^-(s,t) = B_\Delta(s,t), \\
&B_\text{\ref{fig: Deltas}}^3(s,t) = 0,
\end{split}
\end{align}
with
\begin{align}
\begin{split} \label{eq: ADelta}
A_\Delta(s,t)
=&\, \frac{2g^2c_{u-d}}{3f_aF_\pi}\Biggl\{
\frac{2z_0}{3m_\Delta^2} \Big(m_\Delta + \big[m_N+2m_\Delta\big]z_0\Big) \Big(s-m_N^2\Big) \\
&+ \frac{1}{s-\mu_\Delta^2} \biggl[
\Big(m_N+m_\Delta\Big) \Big(\frac{1}{2}\big[m_a^2+M_\pi^2-t\big] - \frac{1}{3}\big[s-m_N^2\big]\Big) \\
&- \frac{1}{6m_\Delta^2} \Big(
\big[m_N+m_\Delta\big] \big[\big(m_a^2+M_\pi^2\big)\big(s-m_N^2\big) \\
&+m_a^2M_\pi^2\big] + m_a^2M_\pi^2 m_\Delta + m_N{\big[s-m_N^2\big]}^2
\Big)
\biggr]
\Biggr\},
\end{split}
\end{align}
and
\begin{align}
\begin{split} \label{eq: BDelta}
B_\Delta(s,t)
=&\, \frac{2g^2c_{u-d}}{3f_aF_\pi}\Biggl\{
-\frac{z_0}{3m_\Delta^2} \Big(m_a^2 + M_\pi^2 + 2\big[s-m_N^2\big]\big[1+z_0\big] \\
&+ 4m_Nm_\Delta\big[1+z_0\big] + 4m_N\big[m_N+m_\Delta\big]z_0\Big) \\
&+ \frac{1}{s-\mu_\Delta^2} \biggl[
\Big(\frac{1}{2}\big[m_a^2+M_\pi^2-t\big] - \frac{1}{6}m_a^2 + \frac{1}{6m_\Delta}\big[m_N+m_\Delta\big]\big[4m_N m_\Delta-M_\pi^2\big]\Big) \\
&- \frac{1}{6m_\Delta^2} \Big(
\big[m_a^2+M_\pi^2+2m_N m_\Delta\big] \big[s-m_N^2\big] + m_a^2\big[m_Nm_\Delta + M_\pi^2\big] + {\big[s-m_N^2\big]}^2
\Big)
\biggr]
\Biggr\}.
\end{split}
\end{align}
Notice that Eqs.~\eqref{eq: ADelta} and \eqref{eq: BDelta} have a pole appearing at c.m. energies around
the $\Delta$ mass. To avoid unnecessary intricacies associated with this, we use a Breit-Wigner propagator
with a complex mass squared,
\begin{equation}
\mu_\Delta^2 = m_\Delta^2 - im_\Delta\Gamma_\Delta,
\end{equation} 
with $\Gamma_\Delta = 117\,\text{MeV}$ the width of the $\Delta$~\cite{ParticleDataGroup:2022pth}. Here, the same comment with respect to the
pole value as already made for the mass applies. A more refined treatment could, e.g., be given by including the $\Delta$ self-energy in the complex
mass scheme~\cite{Hacker:2005fh}, but that is not required here. For the contributions from the exchange
of the $\Delta$ in the crossed channel, Fig.~\ref{fig: Deltau}, analogous relations as the ones shown
in Eq.~\eqref{eq: crossing} hold.

Let us consider now the exchange of the $N^*$ resonance. The Lagrangian for the $N^*\pi{N}$ and $N^*aN$
interactions is~\cite{Bernard:1995dp,Beane:2002ud,Borasoy:2006fk, Djukanovic:2009gt} 
\begin{equation} \label{eq: Roper-pion-nucleon Lagrangian}
\mathcal{L}_{N^*\pi{N}} = \frac{\sqrt{R}}{2}\bar{\Psi}_{N^*}\left\{
\frac{g_A}{2}{u\!\!\!/}\gamma_5 + \frac{g_0^i}{2}{u\!\!\!/}_i\gamma_5
\right\}\Psi_N + \text{h.c.},
\end{equation}
with $\Psi_{N^*}$ the isodoublet Dirac field describing the Roper and $\sqrt{R} = 0.79$ determined
in Ref.~\cite{Meissner:1999vr}.

\begin{figure}
\centering
\subfloat[]{
\includegraphics[width=0.45\textwidth]{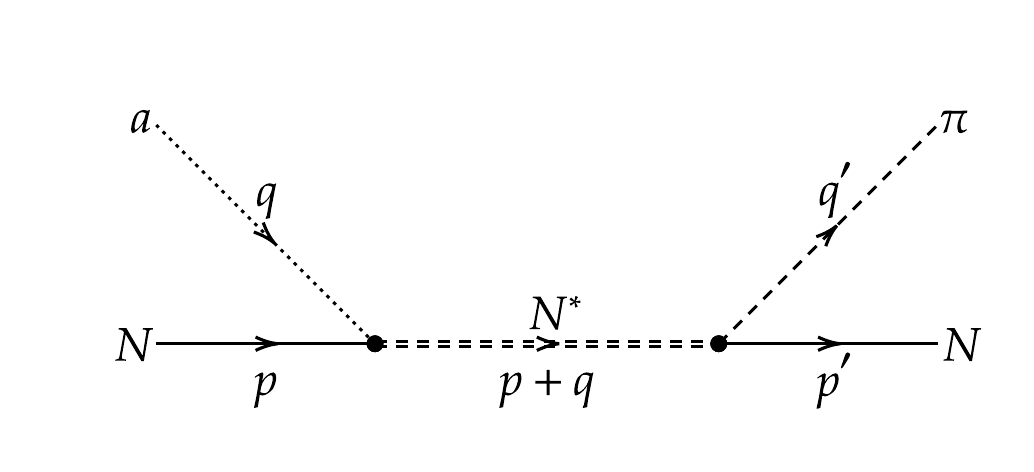}
\label{fig: Ropers}}
\subfloat[]{
\includegraphics[width=0.45\textwidth]{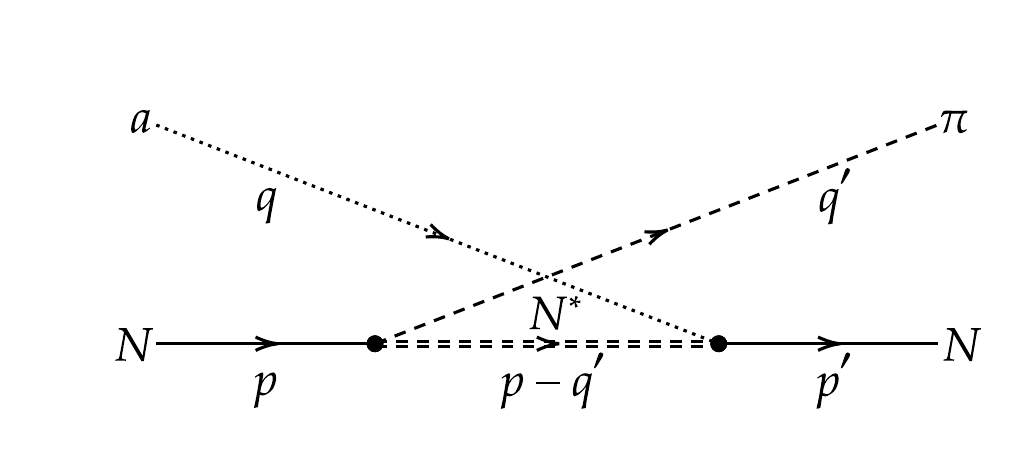}
\label{fig: Roperu}}
\caption{Diagrams  for $aN\to \pi N$ with the (a) $s$-channel and (b) $u$-channel exchange of the $N^*$ resonance.}
\label{fig: Roper}
\end{figure}

The resulting contributions from the Roper-mediated diagrams, Figs.~\ref{fig: Ropers} and \ref{fig: Roperu},
are similar to those from the nucleon-mediated diagrams, see Eqs.~\eqref{eq: 1b} and \eqref{eq: crossing},
with the only difference being the need to replace $A_N$ and $B_N$ with $A_{N^*}$ and $B_{N^*}$, respectively,
\begin{align}
\begin{split}
A_{N^*}(s,t) &= \frac{Rg_A(m_N+m_{N^*})}{16f_aF_\pi}\frac{s-m_N^2}{s-\mu_{N^*}^2}, \\
B_{N^*}(s,t) &= -\frac{Rg_A}{16f_aF_\pi}\left(\frac{2m_N^2+2m_Nm_{N^*}}{s-\mu_{N^*}^2} +\frac{s-m_N^2}{s-\mu_{N^*}^2}\right),
\end{split}
\end{align}
where again we used a complex mass squared in the propagator,
\begin{equation}
\mu_{N^*}^2 = m_{N^*}^2 - im_{N^*}\Gamma_{N^*},
\end{equation} 
with $m_{N^*} = 1440\,\text{MeV}$ and $\Gamma_{N^*} = 350\,\text{MeV}$ the Breit-Wigner mass and width of the Roper resonance~\cite{ParticleDataGroup:2022pth}.
We are well aware that such a simple parametrization does not quite represent the dynamics of the
Roper in pion-nucleon scattering, see, e.g., Ref.~\cite{Gegelia:2016xcw} for a more refined treatment,
but given the exploratory nature of our investigation, it should suffice to estimate the corresponding contribution
to pion axioproduction for c.m. energies below $1.5\,\text{GeV}$.

\subsection{Pion rescattering} \label{sec: rescatter}
\begin{figure}
\centering
\includegraphics[width=0.45\textwidth]{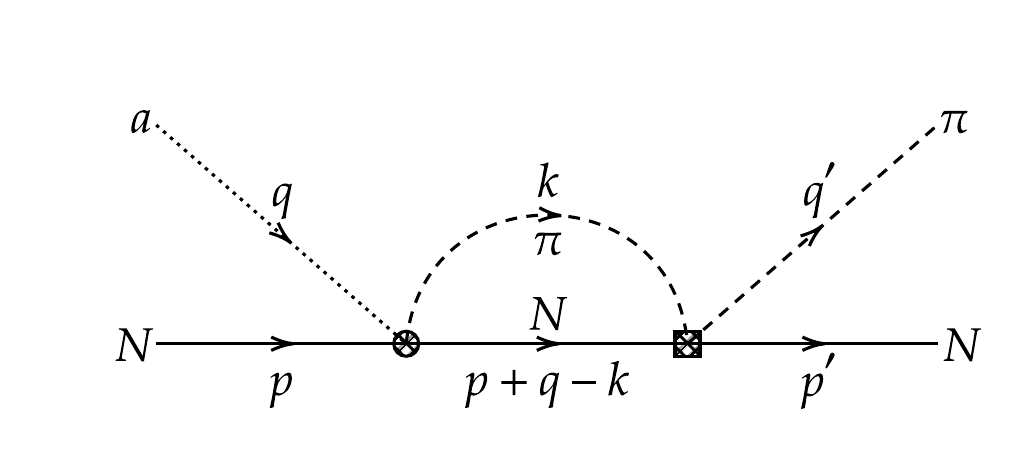}
\caption{The pion rescattering loop diagram for $aN\to \pi N$.}
\label{fig: rescatter}
\end{figure}

The pion rescattering loop diagram is depicted in Fig.~\ref{fig: rescatter}. The resulting contributions
to the partial wave amplitudes $T_{an \to \pi{N}}^{Il\pm}$ ($I$ denotes the isospin of the final $\pi{N}$ system)
can be approximated by
\begin{equation}
T_{an \to \pi{N}}^{Il\pm, {\rm rescatt.}}(s) \approx T_{an \to \pi{N}}^{Il\pm, {\rm tree}}(s)
 \times g(s) \times T_{\pi{N} \to \pi{N}}^{Il\pm}(s).
\end{equation}
The first factor, $T_{an \to \pi{N}}^{Il\pm, {\rm tree}}(s)$, corresponds to the left
vertex which leads to the axion-pion conversion and, therefore, basically comprises the contributions
from Fig.~\ref{fig: contact and nucleon}. The second factor, $g(s)$, is the usual two-point loop function
involving one pion and one nucleon:
\begin{align}
\begin{split}
g(s) &= \frac{1}{16\pi^2} \left\{
\tilde{a}(\mu) + \log{\left(\frac{M_\pi^2}{\mu^2}\right)} - x_+\log{\left(\frac{x_+ - 1}{x_+}\right)}
- x_-\log{\left(\frac{x_- - 1}{x_-}\right)}
\right\}, \\
x_\pm &= \frac{s+m_N^2-M_\pi^2}{2s} \pm \frac{1}{2s}\sqrt{{\left(s+M_\pi^2-m_N^2\right)}^2-4s\left(M_\pi^2-i0^+\right)},
\end{split}
\end{align}
where we fix the regularization scale at $\mu = m_N$ and take the subtraction constant $\tilde{a} = -0.84$ as in
Ref.~\cite{Meissner:1999vr}. Finally, the last factor, $T_{\pi{N} \to \pi{N}}^{Il\pm}(s)$, reflects the effect of
the right vertex which leads to the rescattering of the pion. Since there have been many
studies of pion-nucleon scattering, we do not repeat the computation of $T_{\pi{N} \to \pi{N}}^{Il\pm}(s)$
here and adopt the results of Ref.~\cite{Meissner:1999vr}, see Eq.~(4.11) therein.

\section{Results} \label{sec: results}
In this section, we show and discuss the results of the total and partial wave cross sections of the $an \to \pi^-p$ process.
The final state of the $an \to \pi^-p$ process consists of two charged particles, which can be more easily detected than neutral particles in most experiments.
The results for the other three processes listed also in Eq.~\eqref{eq: T^{+/-/3}} are provided in Appendix~\ref{sec: additional results}.
As advocated in Sec.~\ref{sec: kinematics}, the total cross section can be approximated by the sum of the first three partial wave cross sections,
\begin{equation}
\sigma \approx \sigma^{S_1} + \sigma^{P_1} + \sigma^{P_3}.
\end{equation}
Each partial wave cross section can be calculated by Eq.~\eqref{eq: partial wave cross section}, and the corresponding partial wave amplitude, $T_{an \to \pi^-p}^{l\pm}$, can be obtained from the calculations presented in the previous section,
\begin{equation} \label{eq: tree + loop}
T_{an \to \pi^-p}^{l\pm} \: \approx \: T_{an \to \pi^-p}^{l\pm, {\rm tree}} + T_{an \to \pi^-p}^{l\pm, {\rm loop}}.
\end{equation}
Here, $T_{an \to \pi^-p}^{l\pm, {\rm tree}}$ denotes the contribution arising from the tree diagrams and can be obtained by using Eqs.~\eqref{eq: T^{+/-/3}} and \eqref{eq: partial wave projection} with the functions $A$ and $B$ given in Secs.~\ref{sec: contact and nucleon} and \ref{sec: Delta and Roper}.
$T_{an \to \pi^-p}^{l\pm, {\rm loop}}$ denotes the contribution originating from the loop diagram and can be expanded in terms of $T_{an \to \pi{N}}^{Il\pm, {\rm rescatt.}}$ given in Sec.~\ref{sec: rescatter} by using the isospin decomposition,
\begin{equation}
|\pi^-p\rangle
= -\sqrt{\frac{2}{3}} {|I=1/2\rangle}_{\pi{N}}
  +\sqrt{\frac{1}{3}} {|I=3/2\rangle}_{\pi{N}}.
\end{equation}

\begin{figure}
\centering
\vspace{-12.em}
\makebox[\textwidth][c]{\includegraphics[width=1.15\textwidth]{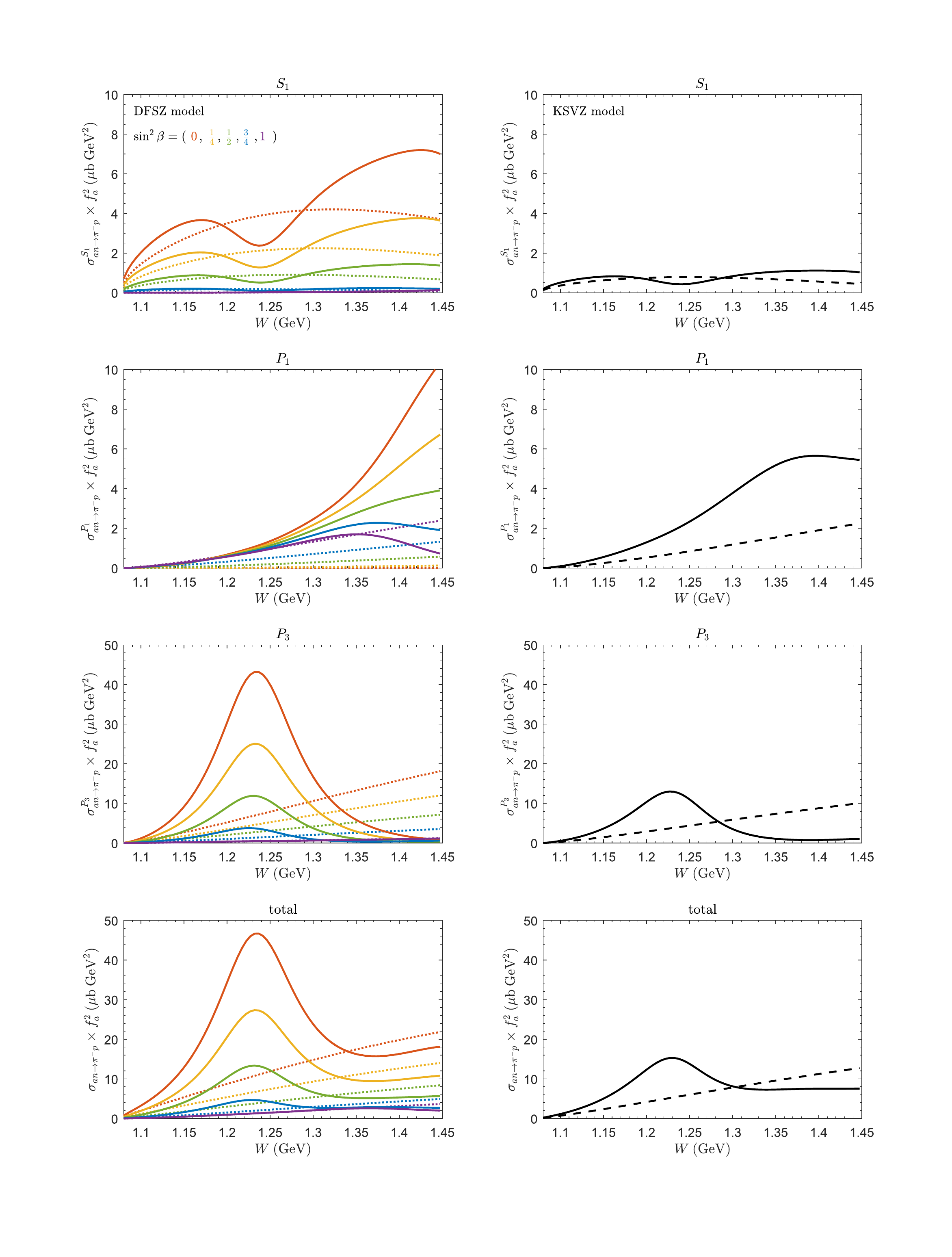}}
\vspace{-7.em}
\caption{
The total and partial wave cross sections of $an \to \pi^-p$ versus the c.m. energy $W$ for the DFSZ axion at different values of $\sin^2{\beta}$ (left panel) and the KSVZ axion (right panel).
In these plots, the solid lines correspond to the cross sections obtained based on Eq.~\eqref{eq: tree + loop}, while the dashed lines correspond to the cross sections obtained considering only the contact and nucleon-mediated diagrams.
In the left panel, the red, orange, green, blue and purple lines correspond to the DFSZ model parameter $\sin^2{\beta} = 0, 1/4, 1/2, 3/4$ and $1$, respectively.
}
\label{fig: channel_4}
\end{figure}

In Fig.~\ref{fig: channel_4}, we show the total as well as the three mentioned partial wave cross sections as functions of the c.m. energy $W$ for the KSVZ and DFSZ models.
For comparison, we also depict by dashed lines the results considering only contributions from the contact and nucleon-mediated diagrams.
The cross sections are multiplied by a factor of $f_a^2$ in order to eliminate the dependence on the unknown axion decay constant.
Additionally, the unknown axion decay constant also implicitly appears in the terms containing the axion mass, but it has negligible practical impact since the axion mass can safely be disregarded within the typical QCD axion window.
Notice that the results of $\sin^2{\beta}$ taking value of $0$ in the DFSZ model is given for illustrative purposes only, as the allowed range for $\tan{\beta}$ due to the perturbative constraints from the heavy quark Yukawa couplings is $[0.25, 170]$~\cite{DiLuzio:2020wdo} corresponding to approximately $\sin^2{\beta} \in [0.06, 1.00]$.

As anticipated, there is indeed an enhancement in the partial wave cross sections of $P_3$ and $P_1$ when $W \sim m_\Delta$ and $W \sim m_{N^*}$ due to the $\Delta$ and $N^*$, respectively.
First, consider the $P_3$ partial wave.
It is evident that the magnitude of the resonance peak decreases as $\sin^2{\beta} \to 1$ in the DFSZ model.
This can be easily understood since the dominant contribution to $T_{an \to \pi^-p}^{P_3}$, arising from the $s$-channel exchange of the $\Delta$, is proportional to $c_{u-d}$ (see Eq.~\eqref{eq: Deltas}), whose absolute value is a linearly decreasing function of $\sin^2{\beta}$ (see Eq.~\eqref{eq: axion couplings}): $\left|c_{u-d}^{\text{DFSZ}}(\sin^2{\beta})\right| = \tfrac{1}{3}\left(1.0116 - \sin^2{\beta}\right)$.
This also explains why the $P_3$ partial wave result of the KSVZ model closely aligns with that of the DFSZ model when $\sin^2{\beta} = \tfrac{1}{2}$, as $c_{u-d}^{\text{DFSZ}}(\sin^2{\beta} = \tfrac{1}{2}) = c_{u-d}^{\text{KSVZ}}$.
We also find that the $P_3$ partial wave results of our work are smaller than the $P_{33}$ partial wave results reported in Ref.~\cite{Vonk:2022tho}.
This discrepancy arises from the fact that it is the isospin eigenstate considered as the initial/final states in that work.
Consequently, the amplitude there takes the form of $X^{3/2} = X^+ - X^-$~\cite{Vonk:2022tho} with $X = A,B$ (see Eq.~\eqref{eq: partial wave projection}), which can be approximated as $\frac{3}{2}X_\Delta$ (see Eq.~\eqref{eq: Deltas}) if one only keeps those contributions from the $s$-channel exchange of the $\Delta$.
In contrast, our analysis considers the physical initial/final states, resulting in an amplitude of $X_{an \to \pi^-p} = -\sqrt{2}(X^- - X^3) \simeq \frac{\sqrt{2}}{2}X_\Delta$.
As a consequence, the peak value of the cross section reported in Ref.~\cite{Vonk:2022tho} ought to be roughly $4.5$ times larger than the one we obtain, basically reflecting the differences.

In the case of $P_1$ partial wave, due to the relatively large decay width of the Roper, its contribution as the intermediate particle in the $s$-channel does not entirely dominate $T_{an \to \pi^-p}^{P_1}$.
Therefore, the dependence of the magnitude of the resonance peak on $\sin^2{\beta}$ in the DFSZ model can no longer be straightforwardly understood through an analysis similar to what we did in the case of $P_3$ partial wave.

We point out again that our total cross section peak in the $\Delta$ region, $\sigma_{an \to \pi^-p} \simeq 50\text{--}1 \, \mu\text{b}\,(\text{GeV}/f_a)^2$ (for $\sin^2{\beta} = 0\text{--}1$), is about a factor of $20$ to $1000$ smaller than the naive estimate given in Ref.~\cite{Carenza:2020cis} because they did not account for the fact that the $\Delta$ contribution is only non-vanishing when isospin symmetry is broken.
In contrast, our results align, within the same order of magnitude, with those reported by Ho et al.~\cite{Ho:2022oaw}.
As an illustration, considering the DFSZ model with $\sin^2{\beta} = 0$, the peak value of the total cross section in the $\Delta$ region in our results is approximately $50 \, \mu\text{b}\,(\text{GeV}/f_a)^2$, while Ref.~\cite{Ho:2022oaw} reported a value around $80 \, \mu\text{b}\,(\text{GeV}/f_a)^2$.
Taking the KSVZ model as another example, both results indicate a peak value of approximately $20 \, \mu\text{b}\,(\text{GeV}/f_a)^2$.

\begin{figure}
\centering
\makebox[\textwidth][c]{\includegraphics[width=0.65\textwidth]{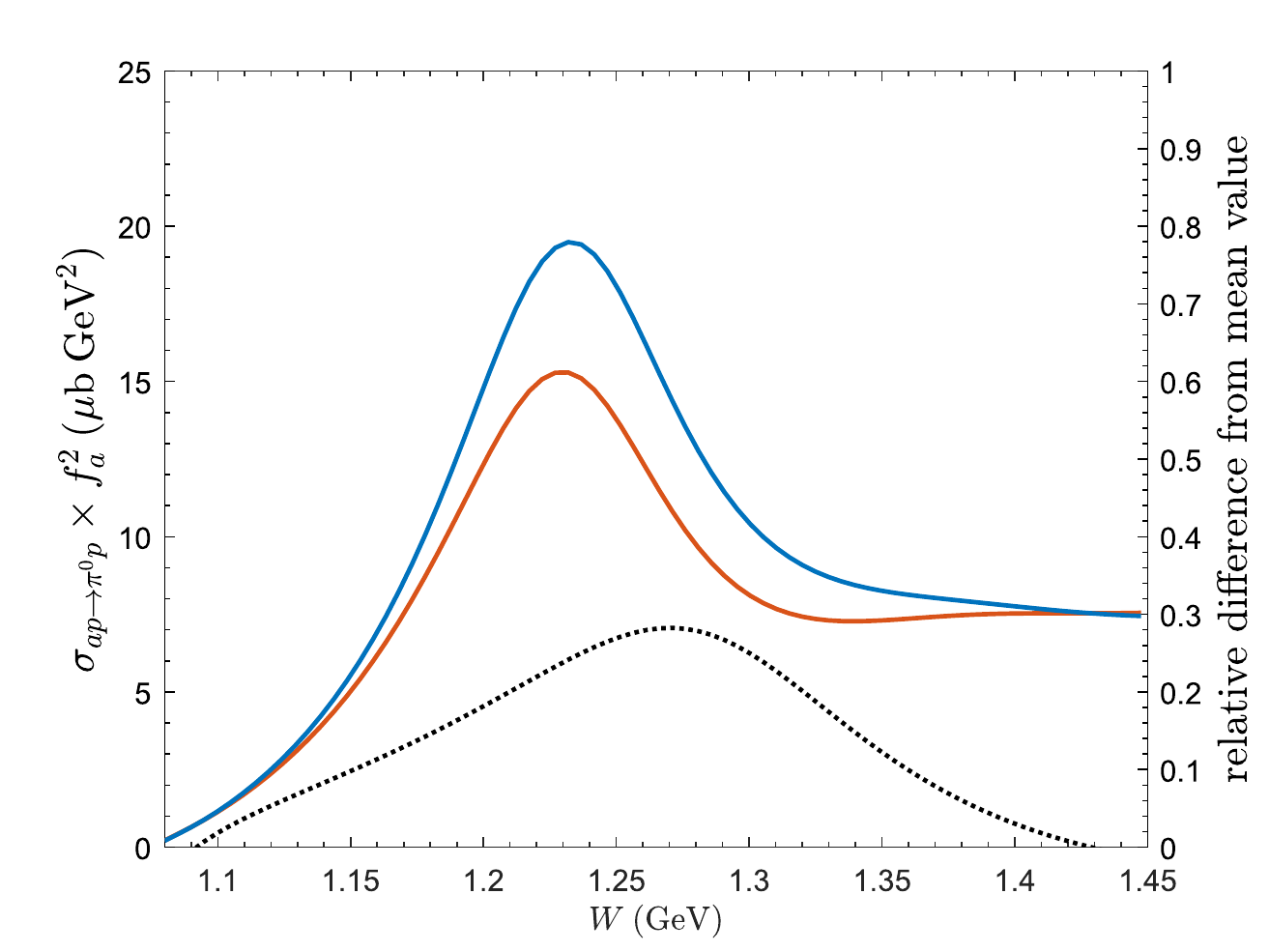}}
\caption{
The total cross sections of the KSVZ model with two different sets of resonance parameters.
The red and blue lines correspond to using the parameters of Fit~2 in Table~1 and Table~2, respectively, in Ref.~\cite{Meissner:1999vr}.
The dashed line corresponds to the relative deviation from the mean value.
}
\label{error_estimate}
\end{figure}

To assess the reliability of our results, we conduct a rough error estimation by performing calculations on the KSVZ axion using two sets of the pertinent $\Delta$ and Roper couplings, selected from distinct fit outcomes presented in Ref.~\cite{Meissner:1999vr}.
We find that the maximal numerical deviation from the mean value is about $30\%$, while the line shapes of the total cross sections remain almost unchanged (up to normalization), see Fig.~\ref{error_estimate}.

\section{Summary}
In this work, we investigated the impact of the $\Delta$ and $N^*(1440)$ as intermediate particles on the cross section of pion axioproduction.
The axions from SNe, that transform into pions in water Cherenkov detectors, can reach energies as high as $500\,\text{MeV}$, making the effects of these resonances non-negligible.
We adopt a chiral Lagrangian framework with the explicit inclusion of resonance fields.
Based on the assumption of resonance saturation, we were able to essentially account for the effects of these resonances by explicitly considering the exchange of them in $s$-channel and $u$-channel,
thereby avoiding the ignorance of LECs related to the higher-order interactions.
The results indicate that an enhanced cross section is indeed present in the region of the $\Delta$ and $N^*(1440)$. However, these effects are drastically reduced compared to the earlier work of Ref.~\cite{Carenza:2020cis}.

Finally, considering the inverse process of pion axioproduction, $\pi{N} \to aN$, as suggested to be the dominant mechanism compared to nucleon bremsstrahlung, $NN \to aNN$, for axion production in SNe~\cite{Carenza:2020cis},
it would be intriguing to explore the influence of these resonances on $\pi{N} \to aN$, which may offer new insights into experimental axion searches.

\begin{acknowledgements}
This work is supported in part by the National Natural Science Foundation of China (NSFC) under Grants No. 12125507, No. 11835015, and No. 12047503;
by the Chinese Academy of Sciences (CAS) under Grant No. YSBR-101;
by NSFC and the Deutsche Forschungsgemeinschaft (DFG) through the funds provided to the Sino-German Collaborative Research Center TRR110 ``Symmetries and the Emergence of Structure in QCD'' (NSFC Grant No. 12070131001, DFG Project-ID 196253076);
by CAS through the President’s International Fellowship Initiative (PIFI) under Grant No. 2018DM0034;
and by the VolkswagenStiftung under Grant No. 93562.
\end{acknowledgements}

\appendix

\section{Additional results} \label{sec: additional results}
\begin{figure}
\centering
\vspace{-12.em}
\makebox[\textwidth][c]{\includegraphics[width=1.15\textwidth]{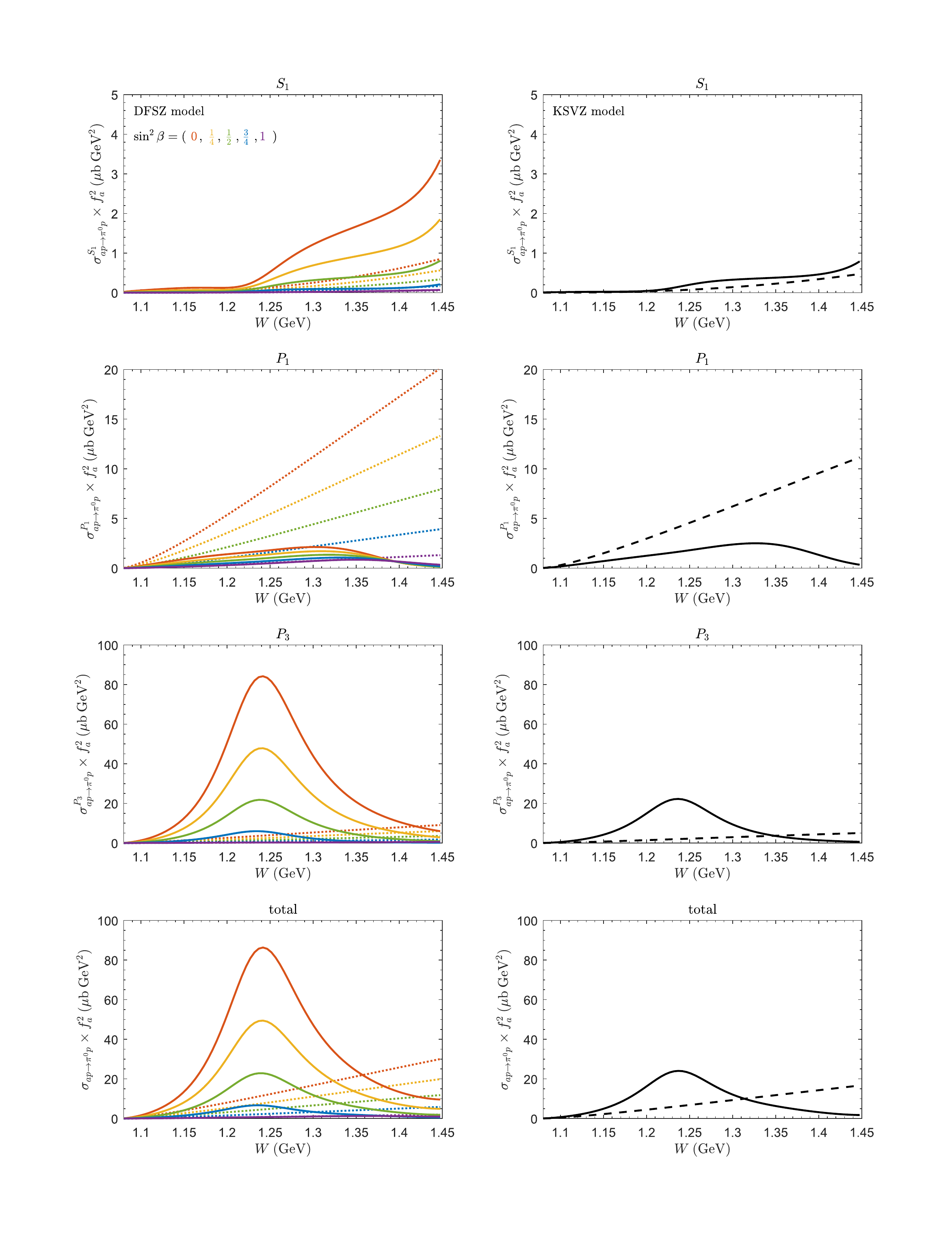}}
\vspace{-7.em}
\caption{The total and partial wave cross sections of $ap\to \pi^0p$ for the DFSZ axion at different values of $\sin^2\beta$ (left panel) and the KSVZ axion (right panel). See the caption of Fig.~\ref{fig: channel_4}.}
\label{fig: channel_1}
\end{figure}

\begin{figure}
\centering
\vspace{-12.em}
\makebox[\textwidth][c]{\includegraphics[width=1.15\textwidth]{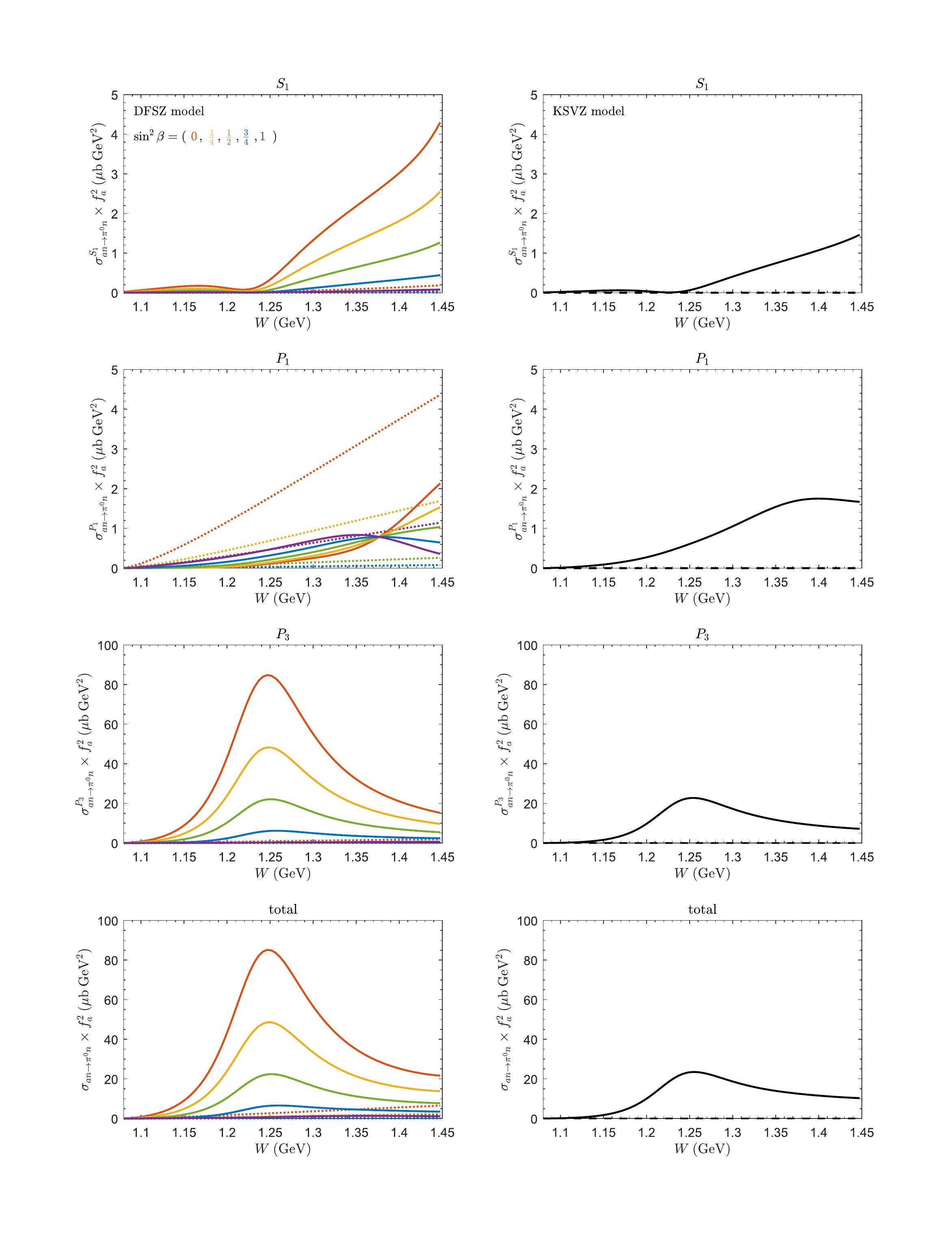}}
\vspace{-7.em}
\caption{The total and partial wave cross sections of $an\to \pi^0n$ for the DFSZ axion at different values of $\sin^2\beta$ (left panel) and the KSVZ axion (right panel). See the caption of Fig.~\ref{fig: channel_4}.}
\label{fig: channel_2}
\end{figure}

\begin{figure}
\centering
\vspace{-12.em}
\makebox[\textwidth][c]{\includegraphics[width=1.15\textwidth]{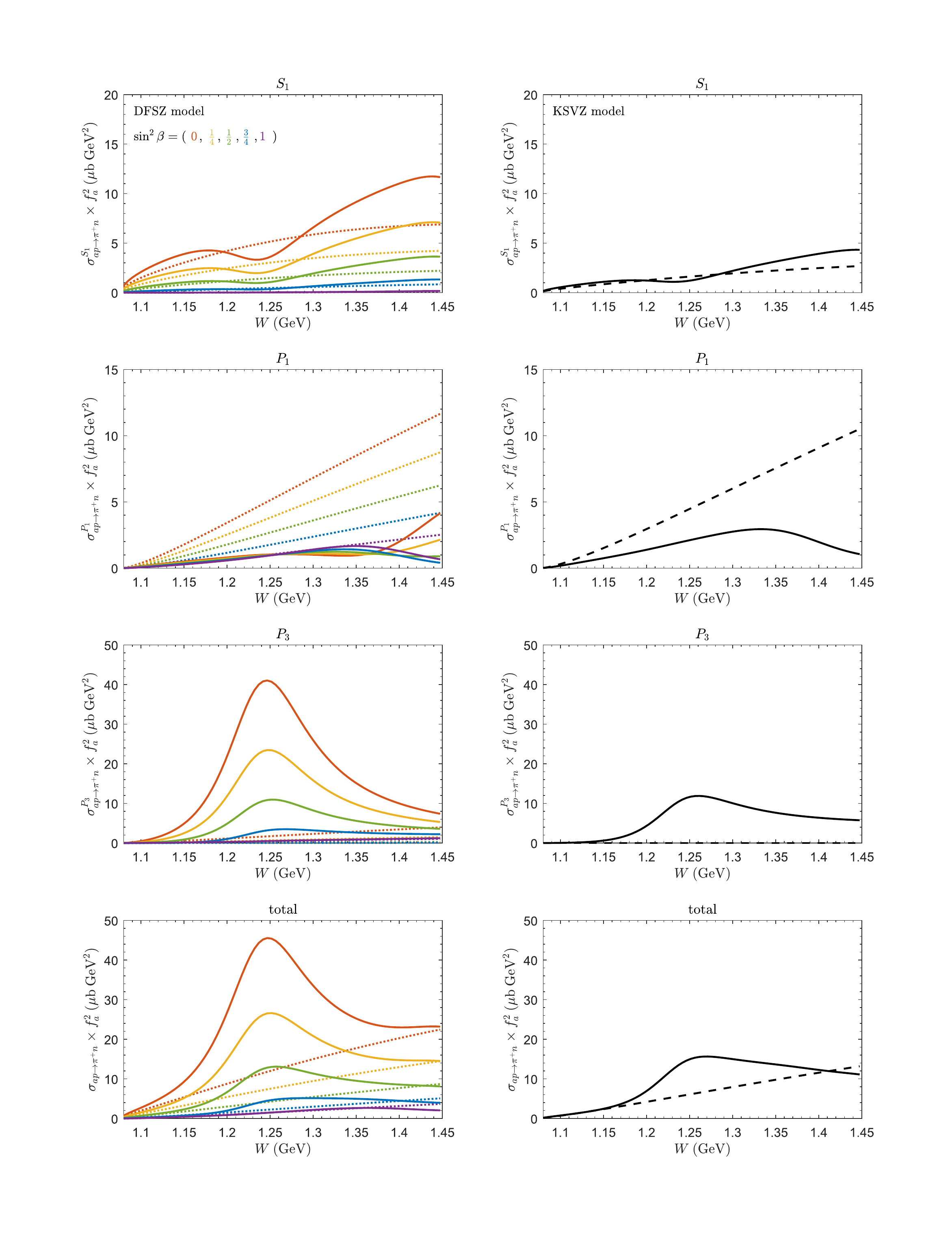}}
\vspace{-7.em}
\caption{The total and partial wave cross sections of $ap\to \pi^+n$ for the DFSZ axion at different values of $\sin^2\beta$ (left panel) and the KSVZ axion (right panel). See the caption of Fig.~\ref{fig: channel_4}.}
\label{fig: channel_3}
\end{figure}

In this appendix, we show the results for three additional processes, $ap \to \pi^0p$, $an \to \pi^0n$ and $ap \to \pi^+n$, which have neutral particles in the final state.
The total and partial wave cross sections of these three processes for the DFSZ and KSVZ axions are shown in Figs.~\ref{fig: channel_1}, \ref{fig: channel_2} and \ref{fig: channel_3}, respectively.

The cross section peak of the $P_3$ partial wave for $ap \to \pi^0p$ and $an \to \pi^0n$ is approximately twice as large as the ones for $ap \to \pi^+n$ and $an \to \pi^-p$.
This observation can be explained by considering that the amplitudes of the former two processes can be approximated as $X_\Delta$, while those of the latter two can be approximated as $\mp\frac{\sqrt{2}}{2}X_\Delta$, if one considers only the $s$-channel $\Delta$ contribution.

\newpage

\bibliography{refs}

\end{document}